\documentclass[iop,usenatbib]{emulateapj}

\usepackage[]{graphicx}
\usepackage{journals}
\usepackage{hyperref}

\shorttitle{MMIRS pipeline}
\shortauthors{Chilingarian et al.}

\begin{document}

%% This is the end of the preamble.  Indicate the beginning of the
%% paper itself with \begin{document}.


\title{Data Reduction Pipeline for the MMT and Magellan Infrared Spectrograph}


\author{Igor Chilingarian\altaffilmark{1,2}}
\email{igor.chilingarian@cfa.harvard.edu}
\author{Yuri Beletsky\altaffilmark{3}}
\author{Sean Moran\altaffilmark{1}}
\author{Warren Brown\altaffilmark{1}}
\author{Brian McLeod\altaffilmark{1}}
\author{Daniel Fabricant\altaffilmark{1}}
\altaffiltext{1}{Smithsonian Astrophysical Observatory, 60 Garden St. MS09, Cambridge, MA 02138
USA}
\altaffiltext{2}{Sternberg Astronomical Institute, Moscow State University,
13 Universitetski prospect, Moscow, 119992, Russia}
\altaffiltext{3}{Las Campanas Observatory, Carnegie Institution of
Washington, Colina el Pino, Casilla 601 La Serena, Chile}



\journalinfo{Publications of the Astronomical Society of the Pacific,
127:000--000}
\submitted{Received 2014 November 05; accepted 2015 January 22; published
2015 April 01}

\begin{abstract}

We describe the new spectroscopic data reduction pipeline for the 
multi-object MMT/Magellan Infrared Spectrograph.  The pipeline is 
implemented in {\sc idl} as a stand-alone package and is publicly available 
in both stable and development versions.  We describe novel algorithms for 
sky subtraction and correction for telluric absorption.  We demonstrate that 
our sky subtraction technique reaches the Poisson limit set by the photon 
statistics.  Our telluric correction uses a hybrid approach by first 
computing a correction function from an observed stellar spectrum, and then 
differentially correcting it using a grid of atmosphere transmission models 
for the target airmass value.  The pipeline provides a sufficient level of 
performance for real time reduction and thus enables data quality control 
during observations.  We reduce an example dataset to demonstrate the high 
data reduction quality.

\end{abstract}

\keywords{techniques: spectroscopic, algorithms, data reduction, sky
subtraction}

%>>>> Include a list of keywords after the abstract

%%%%%%%%%%%%%%%%%%%%%%%%%%%%%%%%%%%%%%%%%%%%%%%%%%%%%%%%%%%%%
\section{INTRODUCTION AND MOTIVATION}
\label{sect:intro}  % \label{} allows reference to this section

The rapid progress in astronomical instrumentation and detector technology 
has resulted in a dramatic increase in the volume and complexity of 
astronomical data.  Data reduction for modern wide-field mosaic imagers and 
multi-object spectrographs has become so computationally intensive and 
complex that individual astronomers can no longer easily process their own 
data.  One response to this problem is dedicated data processing and 
analysis centers, such as TERAPIX \citep{Bertin01}.  Another response is 
user-friendly, professionally maintained data reduction pipelines.

Multi-object spectroscopic observations are among the most difficult types 
of astronomical data to reduce and analyze.  Slit mask spectrographs provide 
a high degree of flexibility in observing science targets over a large field 
of view, but produce datasets with very complex geometric properties that 
depend on slit configuration and optical distortions.

The MMT/Magellan Infrared Spectrograph (MMIRS) is a slit mask based 
cryogenic imaging spectrograph that can be used at the f/5 foci of the 6.5~m 
MMT or the Magellan Clay telescopes.  Its optical and mechanical layouts, 
mode of operations, and control software are described in detail in 
\citet{McLeod+12}.  MMIRS operates in the imaging, longslit and multi-object 
spectroscopy (MOS) modes.  The MOS mode is implemented by installing laser 
cut slit masks into the focal plane.  Slit masks are designed and 
manufactured a few weeks prior to scheduled observations.

Spectroscopic data reduction of near-infrared ($1.0 < \lambda < 2.5 \mu$m)
ground based observations is a non-trivial task that is significantly more
challenging than optical data reduction.  Standard general purpose software
suites used in optical astronomy, such as {\sc noao iraf}, cannot be
used as they do not take into account specific features of the instrument or
its near infrared detector.

Modern NIR detectors still stay about a decade behind optical CCDs in terms
of read-out noise, dark current, and non-linearity of the response.  Also,
they all possess substantial color dependent variations in the
pixel-to-pixel sensitivity reaching 40~per~cent for HAWAII-2 detectors.  An
average number of unrecoverable hot or cold pixels is an order of several
hundreds to tens of thousands on a single 2048$\times$2048 science grade NIR
detector compared to dozens to a few hundreds on a modern optical CCD chip. 
Therefore, specific detector related calibrations are needed in the NIR and
special techniques such as multiple or continuous detector read-outs are
often used \citep{VCR04}.

Another family of difficulties arises from specific night sky properties in
the NIR domain.  Hydroxyl ($OH$) airglow lines dominating the NIR sky
emission in the $H$ and $K$ bands are hundreds of times brighter than
optical $OH$ lines in the $I$ band.  They also possess high variations in
time (minutes) and space (degrees on the sky).  The NIR water vapor telluric
absorption bands cause 100~per~cent absorption (i.e.  are completely opaque)
at some wavelengths and absorb tens of per~cent of the light over extended
wavelength intervals.  Therefore, the sky subtraction and telluric
correction become the most difficult steps of the NIR data reduction. 
Approximate techinques can work well in the optical domain and produce
reasonable results, however, in the NIR where all features are orders of
magnitude stronger, and emission airglow lines are much more numerous, a
different approaches to both observational strategy and data reduction are
needed.

The ``classical'' optical sky subtraction, when the sky spectrum is created
by averaging ``empty'' regions of the slit below and above the target after
linearizing and rectifying a spectral frame and then subtracted at every
slit position, leads to significant artefacts, if applied to NIR data,
making further data analysis impossible.  The advanced sky subtraction
technique by \citet{Kelson03} operates on non-reampled frames and provides
nearly Poisson limited sky subtraction quality for optical long-slit
spectra.  However, because of flat field imperfections and color dependent
pixel-to-pixel sensitivity variations in NIR detectors, this technique does
not provide the desired sky subtraction quality.  Another obstacle specific
to multi-slit datasets is the short length of each individual slit that
affects sky model quality in the \citet{Kelson03} algorithm.

%Given the complexity of MMIRS observing modes, complex algorithms are needed 
%to reduce and analyze the data.  

At present, there are several existing data reduction solutions for optical
multi-object spectrographs using slit masks.  The Gemini Multi-Object
Spectrograph data reduction software is an {\sc iraf} based package
distributed by the Gemini
observatory\footnote{\url{http://www.gemini.edu/sciops/data-and-results/processing-software}}. 
It implements basic data reduction techniques used in optical spectroscopy
and a ``classical'' sky subtraction.  Another two software solutions which
implement the \citet{Kelson03} sky subtraction technique are: a stand-alone
COSMOS package developed and distributed by Carnegie
Observatories\footnote{http://code.obs.carnegiescience.edu/cosmos} that is
used for as a data reduction solution for the IMACS spectrograph operated at
the 6.5-m Magellan telescope, and a Keck DEIMOS data reduction pipeline
\citep{Newman+13}, an {\sc idl} package based on the data reduction suite
from the Sloan Digital Sky Survey project {\sc spec2d}.  There is also the
{\sc spark} package \citep{Davies+13} for the data reduction from the KMOS
multiple integral field unit NIR spectrograph operated at the European
Southern Observatory Very Large Telescope.  However, this instrument has a
fixed geometry and the specific observational strategy is used, therefore
the pipeline cannot easily be modified to reduce slit mask data.

Our goal is to design and implement an efficient and flexible system to
reduce multi-slit and longslit spectrosopic data collected with MMIRS.  Our
intention is make this a stand-alone package relying as little as possible
on third-party software.  The pipeline should be easy to install and use by
users not connected to the instrument team and also provide a sufficient
level of performance for real time reduction to enable data quality control
during observations.

Our approach is to control the pipeline with text files having the 
FITS-header like format, i.e.  keyword--value pairs, easy to create and edit 
and similar to {\sc noao iraf} packages.  We pay special attention to the 
quality of sky subtraction, as it is perhaps the single most important step 
required for the successful analysis and interpretation of near-infrared 
(NIR) datasets.  We also develop a technique for the telluric correction of 
NIR spectra.  Our approach requires only one telluric star observation and 
then computes differential corrections based on a grid of pre-computed 
atmosphere transmission models.

Data reduction for MOS and long-slit data in the optical and NIR share 
common algorithms for some reduction steps.  The overall layout of a NIR 
spectral pipeline for MOS data resembles that of pipelines for optical MOS 
and integral-field unit (IFU) spectrographs.  Here we partially adopt the 
structure of the data reduction package for the Multi-Pupil Fiber 
Spectrograph (MPFS, \citealp{ADM01}) at the Russian 6-m telescope described 
in \citet{CPSA07}.  This data reduction package was later transformed into a 
universal data reduction pipeline for integral-field unit (IFU) 
spectrographs and then used to reduce data from ESO VIMOS \citep{CdRB09}, 
Calar-Alto PMAS \citep{CB10}, and Gemini GMOS-IFU.  Some blocks of the code 
in the current MMIRS data reduction pipeline (e.g.  certain diagnostic 
plotting routines, the quality assessment for the wavelength calibration) 
and the task control file structure are based on ideas from that IFU data 
reduction package.

The paper is organized as follows. In Section~2 we briefly present current
MMIRS capabilities including the upgrade made in 2014.  In Section~3 we
describe the format of raw data produced by the MMIRS data acquisition
software and describe the observing protocols for spectoscopic observations. 
In Section~4 we describe the data reduction pipeline, specific algorithms we
developed and implemented, the output data format In Section~5 we discuss
the quality of the sky subtraction that we achieve using the pipeline.  We
summarize in Section~6.  The appendices describe associated calibration
datasets, provide a list of third-party routines used in the pipeline
code, and give an example of the pipeline control file.

\section{MMIRS CAPABILITIES}
\label{sect:capabilities}

\begin{figure*}
\includegraphics[width=\hsize]{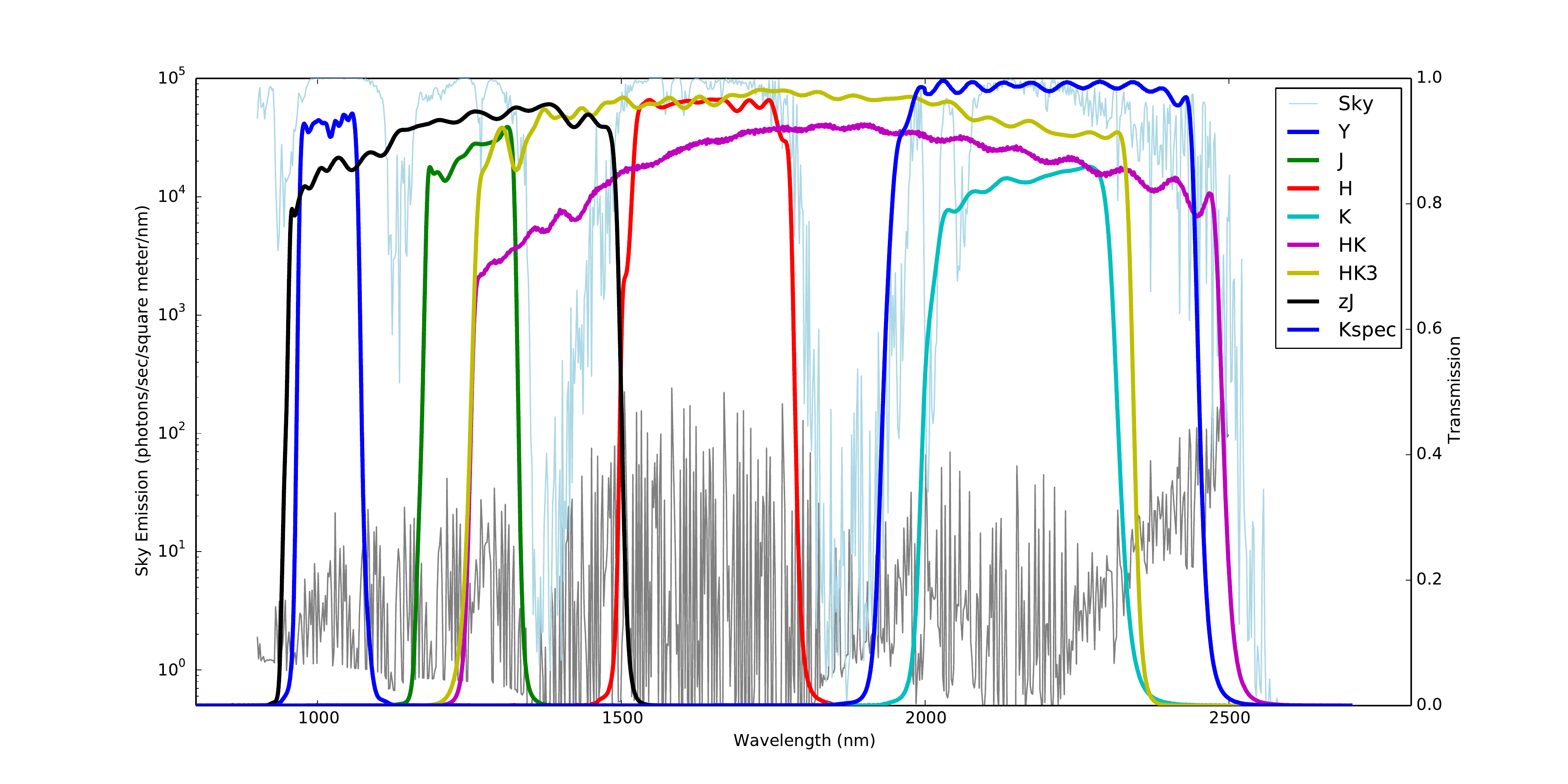}
\caption{Transmission curves of filters available in MMIRS as of October 2014
are shown over the troughput spectrum of the earth atmosphere (light blue)
and the airglow emission spectrum (grey).
\label{irsky}}
\end{figure*}

MMIRS is capable of slit-mask multi-object spectroscopy over a 
$4\times6.9$~arcmin field, long-slit spectroscopy over a 6.9 arcmin slit, 
and direct imaging in the $Y$, $J$, $H$, $K$ bands.  A set of grisms 
provides low- to intermediate-spectral resolution ($R = 1100 \dots 3000$) 
spectroscopy, and are used in a combination with bandpass selection filters 
($zJ$, $HK$, and $Kspec$ for the $J$, $HK$, and $K3000$ grisms respectively) 
or a standard $H$-band filter for the $H3000$ grims (see Fig.~\ref{irsky}).  
\citet{McLeod+12} provide a detailed description of the MMIRS spectral and 
imaging capabilities, here we outline the updates since 2012.

In the fall of 2014 we preformed a number of upgrades in the spectrograph,
first available to observers during the 2014B run on the 6.5-m Magellan
Clay telescope.
\begin{itemize}
\item We replaced the HAWAII2 detector with HAWAII2-RG. The new detector
provides 10 times lower dark current, increased sensitivity at shorter
wavelength (0.8--1.3~$\mu$m), lower read-noise level (11~$e-$), and much
lower pixel-to-pixel sensitivity variations. The array is read through 32 
amplifiers in 1.475~sec.
\item We installed two new virtual phase holographic grisms, $H3000$ and 
$K3000$, that provide $R\approx3000$ spectral resolution $R\approx3000$ in 
the $H$ and $K$ bands, respectively.
\item We installed a new $HK$ spectral bandpass filter called $HK3$.  
Compared to the old $HK$ filter, the new $HK3$ filter has higher overall 
throughput and a shorter wavelength cut-off at 2.33~$\mu$m that reduces by a 
factor of 4 the scattered light originating from the thermal glow of the 
gate valve.
\item We installed the $Kspec$ high throughput bandpass filter, with a 
cut-off at 2.45~$\mu$m designed specifically for $K$-band spectroscopy with 
the newly installed $K3000$ holographic grism. 
\end{itemize}

\section{MMIRS DATA FORMAT AND OBSERVING PROTOCOL}
\label{sect:format}

\subsection{Raw data format} 
MMIRS is equipped with a HAWAII-2RG (formerly, HAWAII-2) detector that is 
read non-destructively ``up-the-ramp'' via 32 amplifyier channels at a 
cadence set by the observer, typically once every 1 or 5 seconds depending 
on the exposure time.  Every read-out is stored as a two-dimensional image 
extension in a FITS file.  The last read-out is put into the first image 
extension in order to facilitate the computation of double correlated frames 
for quick-look purposes by subtracting the two first consequent extensions 
from each other.  The final multi-extension FITS files are stored in the 
MMIRS raw data archive and used as the input product for the data reduction 
pipeline.

It is important to notice that the multi-extention FITS data format is the
same for every type of MMIRS observing mode: imaging, long-slit or
multi-object spectroscopy. However, MOS raw data also contain an additional
text file describing the slit mask layout (mask definition file hereafter).

\begin{figure*}
\includegraphics[width=\hsize]{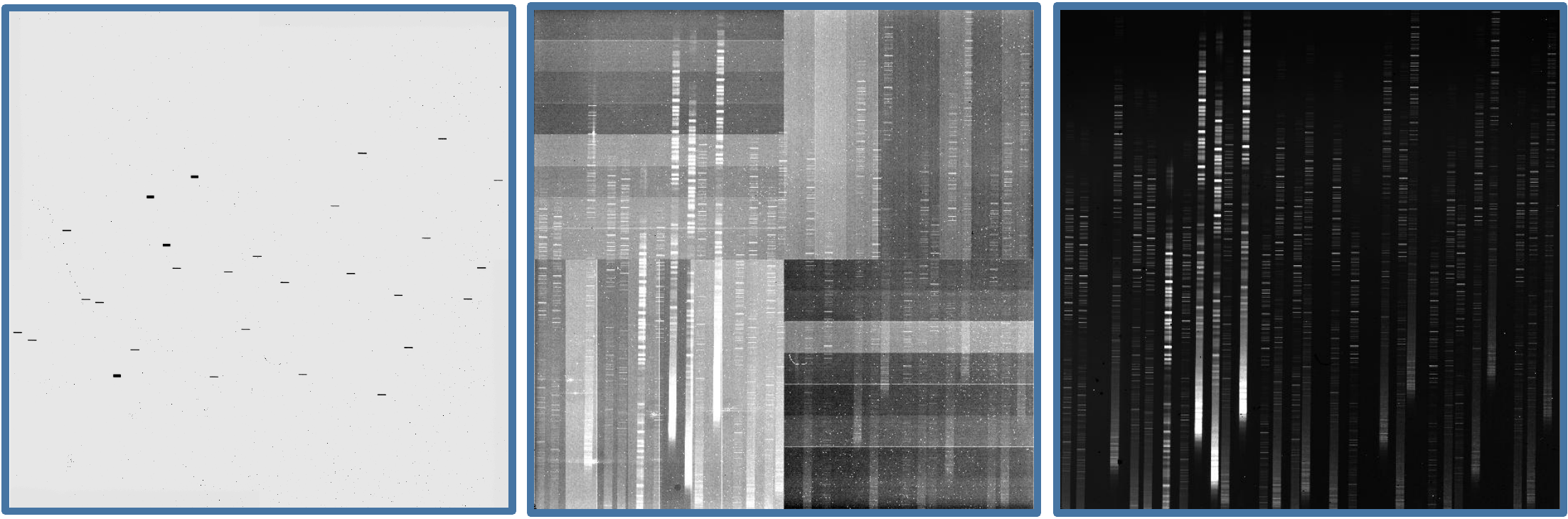}
\caption{A slit mask image (left panel), 
a raw readout of its spectrum from the HAWAII-2 detector (central panel)
and the result of the up-the-ramp fitting and dark
subtraction (right panel). Individual spectra are dispersed in the vertical
direction. Four alignment star boxes visible as ``bold'' slits on the left
panel correspond to four brighter spectra in the left part of the central 
and right panels.
\label{figrawspec}}
\end{figure*}

The first step of the data reduction process is thus, for each pixel, to fit
a smooth monotonic function to the values obtained in that pixel at every
read-out (see Fig.~\ref{figrawspec}).  This approach, usually referred to as
``up-the-ramp fitting'' offers a number of advantages over a standard double
correlated frame, that is a difference between the first and final read-outs. 
Up-the-ramp fitting (a) reduces the read noise because the same pixel is
read multiple times; (b) takes into account the detector non-linear
response; (c) recovers fluxes from pixels saturated in the middle of an
exposure after a few read-outs; and (d) recovers fluxes from pixels affected
by cosmic ray hits in the middle of exposure.

In the current implementation of the MMIRS data reduction pipeline, we
perform the non-linearity correction using pre-computed polynomial
approximations to the non-linear detector response derived from calibration
data.  This approach allows us to obtain fluxes using a linear function in
the up-the-ramp fitting procedure.

\subsection{Spectroscopic observing protocol}

To assure that spectral data collected with MMIRS can be reduced by our
pipeline, observers should use the following observing protocol.

The pipeline works on pairs of spectra spatially dithered along the slit 
taken one after another.  Therefore, observer should use a \emph{dithering 
pattern} with an even number of positions chosen in such a way that the 
distance between positions in every pair stays constant.  The distance must 
be a multiple of 0.2~arcsec (one pixel on the detector) and should be at 
least 3 times as large as the image quality FWHM to avoid contamination of 
spectra on difference frames by their dithering counterparts.  The 
recommended dithering pattern for 6~arcsec long slits, the most commonly 
used slit length in MOS masks, is $+$1.6 $-$1.2, $+$1.2 $-$1.6~arcsec.  For 
longslit observations, this pattern depends on the spatial extent of an 
observed source and may be as wide as $\pm$120~arcsec for large galaxies and 
nebulae.

We strongly recommend taking \emph{night time calibrations} because the
wheel positioning inside the spectrograph is not 100~per~cent repeatable. 
Observers should acquire spectral flat fields, therefore, before any action
that requires moving a grism or a slit or MOS mask (e.g. checking the slit
or MOS mask alignment).  For faint targets and long exposure times (180sec
or longer), we use airglow \emph{OH} lines for the wavelength calibration
(see below), however, we still recommend to take arc frames (\emph{Ar}) as a
part of the calibration plan after taking flat fields. This takes less than a
minute and will provide a backup option if the wavelength calibration using
\emph{OH} lines fails.

We recommend checking the \emph{slit or MOS mask alignment} by removing the
grism and taking a slit image every couple of hours.

We recommend observing \emph{telluric standard stars} in order to remove the 
atmosphertic absorption pattern from spectra and perform relative flux 
calibration (see below).  We strongly recommend using only \emph{A0V} stars. 
The current algorithm in the MMIRS pipeline uses one telluric standard 
observation and scales it to the airmass of an observation using atmosphere 
transmission models.  However, it does not take into account possible 
changes in the water vapor level.  Therefore, we recommend observing a 
telluric standard at least once per field, and no less than 3 times per 
night if one field is observed the whole night.  If a large airmass range is 
covered during the observation of one field, we recommend observing the 
telluric standard around the lowest and the highest airmass, so that there 
will be a choice during the data reduction. In the MOS mode, we recommend 
observing a telluric standard through 5 slits, central, top, bottom, left, 
and right, in order to achieve the complete wavelength coverage and high 
signal-to-noise ratios.  Telluric standards should be chosen to be faint 
enough to avoid saturation, but bright enough to achieve peak counts of 
several to twenty thousands in every individual spectrum. For the longslit 
mode, a telluric standard is observed using the same procedure as a science 
target with a dithering including two positions, for example $+$10 and 
$-$10~arcsec.

The \emph{mask design} must comply with the following guidelines: (1) all 
science target slits should have equal widths, because the sky subtraction 
technique relies on this assumption; (2) alignment star boxes should have 
similar slit lengths to the science target slits, because the pipeline uses 
the alignment star spectra as references when co-adding individual 
observations.  If the alignment stars are dithered out of their boxes, the 
co-adding procedure will fail and instead assume that the objects are 
perfectly centered in their slits, assuming the commanded offsets.

\section{DATA REDUCTION PIPELINE}
\label{sect:pipeline}

The MMIRS data reduction pipeline is a stand-alone package implemented in
{\sc idl}.  It can be used to reduce both long-slit and multi-object
spectroscopic MMIRS observations in a fully automatic way.  Some of the
pipeline blocks may be used to reduce imaging data as well.  With minor
modifications, the pipeline can be used to reduce data from other optical
and NIR multi-object spectrographs.

The MMIRS data reduction pipeline is distributed under the {\sc gpl}
license.  A stable version of the pipeline code (currently v.~1.0) is
distributed as an archive from the web-site of the Telescope Data
Center\footnote{\url{http://tdc-www.harvard.edu/software/mmirs_pipeline.html}}.  The current development
version of the code is available from the public {\sc git} repository at
{\sc
bitbucket}\footnote{\url{https://bitbucket.org/chil_sai/mmirs-pipeline}}.
The MMIRS pipeline depends on the {\sc astrolib} package distributied by 
the NASA Goddard Space Flight Center.

The pipeline is controlled by a task control file having a format similar to
FITS headers, i.e. a set of keyword -- value pairs where the keyword length
is limited to 8 characters and the total line length is limited to 80
characters. This file can be created either in a text editor, or generated by an
automated system that parses observing logs, and then modified if necessary.

This file can be edited in order to perform a specific data reduction step,
but by default the data are reduced completely until the sky subtraction
step.  We adopted the task control file approach from the IFU data reduction
package mentioned earlier.

At present, all of the metadata associated with existing MMIRS observations
are stored in a relational database using PostgreSQL database management
system.  We have a script implemented in {\sc perl} that performs searches
in this database based on object coordinates or name. This script then
finds the best suitable calibration frames including telluric standard
stars.  Finally, the script generates pipeline task control files for
the entire observing sequence of a given target, which may include hundreds
of frames.  Each such file normally includes information for two spatially
dithered exposures.

The pipeline supports long-slit and multi-object spectroscopic observations
through 1 to 12 pixels (0.2 to 2.4~arcsec) wide slits/slitlets. We list the
grism and filter combinations currently supported by the pipeline in
Table~\ref{tabgrism}.

\begin{table}
\caption{Grisms and filter combinations available as of October 2014 and
their support status in the MMIRS pipeline v~1.0.\label{tabgrism}}
\begin{center}
\begin{tabular}{llllc}
Grism & Filter & Sp.Res ($R$) & $\lambda$, $\mu$m & Supported \\
\hline
\hline
$J$ & $J$ & 2200 & $1.15-1.35$ & yes \\
$J$ & $zJ$ & 2200 & $0.95-1.50$ & yes \\
$H$ & $H$ & 2300 & $1.50-1.80$ & yes \\
$H$ & $HK$ & 2300 & $1.25-2.15$ & yes \\
$HK$ & $HK/HK3$ & 1200 & $1.25-2.45$ & yes \\
$H3000$ & $H$ & 3000 & $1.50-1.80$ & planned \\
$K3000$ & $Kspec$ & 3000 & $1.90-2.45$ & planned \\
$HK$ & $zJ$ & 2400 & $0.95-1.50$ & no \\
$HK$ & $Y$ & 3000 & $0.95-1.10$ & no \\
\hline
\end{tabular}
\end{center}
\end{table}

\subsection{Primary data reduction}

The primary data reduction includes the following two steps (see Fig.\ 
\ref{figrawspec}):
\begin{itemize}

\item \emph{Applying the non-linearity correction and fitting the
up-the-ramp slope} in every pixel.  The result of this step is a
two-dimensional image.  We perform the non-linearity correction using a set
of pre-computed polynomial functions (7th to 9th orders) available as a part
of the MMIRS pipeline calibration data package, see below.  At this stage,
any unrecoverable saturated pixels are also masked out.  The counts,
corrected for the non-linear detector response, are then fit with a linear
function.  The slope of the line is the flux rate in that pixel.  Because
this step is computationally intensive, and because the code does not
require any input parameters, it is executed only when a dataset is reduced
for the first time.  Unlike further steps of the data reduction which are
repeated as needed, the code checks for the presence of pre-processed 2D
frames to avoid repeating this step.  If it is necessary to re-run this data
reduction step, the pre-processed 2D frames should be deleted from the input
directory.

\item \emph{Dark subtraction} is the next step, and it is done to the 
up-the-ramp processed 2D frames.  We average several dark frames and 
subtract them from science and calibration frames.  Dark frames are obtained 
after a night of observations by a dark script for every up-the-ramp cadence 
and exposure time combination.  Since we use different up-the-ramp cadence 
and exposure time settings for different types of data, we use different 
dark frames for processing science exposures, flat fields, comparison 
spectra and telluric standards. 
\end{itemize}

These steps are detector specific and they rely on the calibration data
distributed with the pipeline code: non-linearity correcting polynomials and
saturation limits. They can be used for a different detector, once
corresponding calibration datasets have been updated. In Fig~\ref{figrawspec}
we provide an example of a raw read-out from the MMIRS detector and the
result of the primary data reduction applied to the same dataset.

\subsection{Spectral tracing, flat fielding, and wavelength calibration}

After the primary data reduction is complete, we subtract pairs of dithered
spectral exposures (difference spectra hereafter) and process them through
the rest of the pipeline steps.  Individual exposures need to be processed
in order to derive flux uncertainties, done by propagating the initial 
photon statistics through to the final data products.

The following data reduction steps are typical for any slit mask
spectrograph.  We provide our own implementations of all algorithmic blocks
with as few dependencies on external {\sc idl} packages as possible.  These
solutions can be used in the data reduction pipeline for any optical or NIR
spectrograph with only minor modifications.

Long-slit spectra are reduced as a degenerated case of MOS fields with only
one ``target'' slit and no ``alignment star'' slits.  At this stage we
generate a mock mask defintion file for a long-slit setup using the same
format as that for real MOS masks.  This allows us to use the same
code and the same data reduction approach for all types of
MMIRS spectral data.

\begin{itemize} 
\item \emph{Spectral tracing} is a mandatory first step of the MOS data reduction. 
The pipeline reads the MOS mask definition file, then uses it to trace the
two-dimensional spectra, and the gaps between them, on a spectral
flat field exposure. This information is stored in a table and used in the
subsequent steps of the data reduction.

\item \emph{Diffuse light modelling and subtraction} is a step that can be
optionally performed for MOS observations.  It is also skipped when dealing
with difference spectra, because the diffuse light component is effectively
cancelled by subtracting individual frames.  Using the spectral tracing
results, we measure count levels in the gaps between slit images in science
exposures, telluric standards, and calibration frames.  Then, a smooth
background model is constructed from those meaurements using low-order
polynomials along the dispersion direction, and basic splines ($b$-splines
hereafter) across the dispersion direction.  This approach models the
diffuse background, but it cannot handle ghost spectra originating from
reflections off the back surface of the mask.  The amplitude of ghost
spectra is about 1~per~cent of the flux, however, and so they do not
seriously affect the rest of the reduction process unless very bright
sources are observed or used as alignment stars.

\item \emph{Mapping the optical distortions} is a step performed using the 
results of the flat field tracing.  We use a low order 2D polynomial 
approximation to describe distortions originating from both the collimator 
and camera optics, sufficient to provide us with a sub-pixel precision at 
the end.  The distortion map accounts for the imperfect positioning of the 
grism, which causes spectra to be not exactly parallel to the detector edge 
(0.2 to 0.7~deg for the current grism setups), and also for the subtle 
variation of the spatial scale along the wavelength direction.  The 
distortion maps turn out to be very stable in time and do not differ 
significantly from one mask to another. Therefore we store a template 
version of the distortion map for every grism, and compare the derived 
distortion map against the appropriate template during the data reduction.  
If the distortion maps differ significantly, which may happen in case of 
non-uniform filling of the MOS mask by slits, we use the values from the 
template version. Hardware maintenance on MMIRS optical and mechanical 
elements in April of 2014 caused the rotation angle of the $HK$ grism 
to change slightly. In order to handle this, we keep two versions of the 
\emph{HK} grism calibration, and the pipeline decides which one to use based 
on the date of observations stored in the FITS header.

\item \emph{Extraction of two-dimensional spectra} is done after tracing and
distortion mapping.  The pipeline reads the MOS mask definition files and
extracts 2D sub-frames corresponding to bounding boxes of individual spectra
from the original frames with no resampling.  This operation
is performed on both individual exposures and subtracted pairs.  If one of
the neighboring spectra enters the bounding box, it is masked out.

\item \emph{Flat fielding} is a very important part of the data reduction 
procedure because pixel-to-pixel sensitivity variations in the {\sc 
hawaii-2} detector reach 35~per~cent and they are dependent on the spectral 
content of the light.  It is also required to handle flux variations along 
slits due to the imperfect mask cutting (or similar defects on long slits 
permanently installed in the spectrograph).  We normalise flux to unity 
using the 5$\times$5 median-averaged frame as a reference in order to 
exclude possible cosmic ray hit or hot pixel residuals.  For MOS data there 
are two options: (a) normalisation to the maximal flux in one of the slits 
and (b) per-slit normalisation.  The internal flat field lamp provides 
imperfect illumination of the detector plane, therefore we recommend using 
the per-slit normalisation.

\item \emph{The wavelength calibration} is done slightly differently for
long-slit and MOS modes.  Depending on the exposure time and the brightness
of the science target (controlled by the {\sc bright} keyword in the
pipeline control file) we use either internal argon arc lamp or atmosphere
airglow \emph{OH} lines to construct the wavelength solution.  By default, the internal
\emph{Ar} lamp is used only in case of short exposures, when the number of
well-exposed airglow lines is insufficient to achieve the desired quality of
the wavelength solution (1/20 of a pixel or better). Also, the wavelength
solution computation procedure will use internal arc frames if the
\emph{OH} based computation fails for any reasons.  However, a pipeline
user can force using internal arcs or \emph{OH} lines with the corresponding
setting in the pipeline control file. The pipeline calibration
data package includes two separate line lists (\emph{Ar} and \emph{OH})
which contain vacuum wavelengths, relative intensities of different lines,
and a priority flag (0 to 3) for the line usage with ``3'' being the highest
priority and ``0'' for the lines to ignore in the wavelength solution.

A model \emph{Ar} or \emph{OH} spectrum is constructed using the information
from the line list and the initial approximation of the wavelength solution
derived once for every grism and stored in one of the pipeline routines. 
Then, an observed spectrum is compared to the model and emission lines are
automatically identified using an iterative procedure with outlier removal. 
This procedure includes a piecewise cross-correlation to find an initial
shift between a spectrum and a model, then it searches for emission lines
and identifies them against the line list.  Finally, the initial wavelength
solution is approximated using a two-dimensional polynomial (2nd or 3rd
order in both directions, along and across dispersion).

This step is applied to the entire frame in the long-slit mode, or to
every extracted slit in the MOS mode.  Alignment
star boxes are not processed at this stage because the spectral lines are
too broad and many of them are blended.

The second step of the wavelength calibration is applied to MOS data only:
we approximate the behaviour of polynomial coefficients in all slits
simultaneously as smooth two-dimensional polynomial functions of the $(X,Y)$
slit position in the mask.  This allows us to handle alignment star boxes by
simply evaluating the resulting polynomials in the corresponding positions
on the mask.  This will produce a useable wavelength solution if the
masks were cut perfectly.  However, since slits always contain some
imperfections on their edges, we need to perform a final adjustment of the
wavelength solution at every position in every slit.  

The last step, which is also applied to long-slit data, creates a ``template
spectrum'' using a science frame assuming that all targets are faint and
airglow lines are prominent and bright (which is true in over 90~per~cent of
cases).  The ``template spectrum'' is created from spectra resampled in
wavelength using the preliminary wavelength solution and taken close to the
center of the frame.  Then the spectra at every position across dispersion
are cross-correlated against the template spectrum and the zero point of the
wavelength solution is adjusted corresponding to the position of the
correlation peak.  This approach may fail if targets are bright.
In this case, the same procedure can be
performed on the internal arc spectrum, however it will not assess possible
systematic offsets of the wavelength solution which may arise from the
optical system not being ideally telecentric.

\end{itemize}

\subsection{Sky subtraction}

Reliable and precise subtraction of night sky emission is one of the
most critical steps in NIR data reduction.  NIR night sky spectrum
contains very bright emission lines (mostly hydroxyl, OH) as well as a
continuum background that change in time and across the field of view. 
Faint targets observed with MMIRS with integration times of many hours are
often hundreds of times fainter than the NIR night sky level.  

In the MMIRS pipeline we use a hybrid approach to subtract the sky emission
using a combination of classical and recently developed techniques.  As
mentioned above, we create difference images that we process through the
pipeline, which is a classical approach to sky subtraction in NIR.

However, difference spectra usually exhibit residual night sky emission
originating from the short-term time variations of OH lines on a
timescale of minutes.  For the typical 300--600~sec duration of science
exposures, these variations may reach a few per~cent (sometimes tens of
per~cent), therefore it is important to tackle the line residuals which can
be positive or negative.

We use the sky subtraction technique proposed by \citet{Kelson03} which we
modified for our multi-slit spectroscopic data.  The main idea of this sky
subtraction technique is to use non-resampled spectral frames and precise
pixel-to-wavelength mapping in order to create an oversampled model of the
night sky spectrum.  This oversampled night sky spectrum is then
parametrized using $b$-splines and evaluated at every position at every
slit.  Using this approach effectively eliminates artefacts originating from
the interpolation of extremely sharp and undersampled emission line profiles
of airglow \emph{OH} lines.

However, the Kelson sky subtraction approach requires significant curvature 
of slit images on the detector frame in order to properly oversample the 
night sky emission lines.  While this is true for the long-slit 
spectroscopic mode of MMIRS, where the curvature exceeds 10 pixels along the 
slit length, in the MOS mode the situation is very different.  For a typical 
slit length of 7--8~arcsec, the curvature is close to zero in the central 
slit images and only 1 or 2 pixels at the edges of the field of view.  
Therefore, we cannot use the standard \citet{Kelson03} technique in the MOS 
mode.

We modify the Kelson algorithm by adding an extra dimension to the sky
spectrum parametrization: the published approach uses the pixel position
along the slit as an extra parameter, while we use 2-dimensional physical
positions of every pixel on the slit mask.  This allows us to take into
account the residual flat fielding errors, both along the dispersion
direction and across the field of view.

After the wavelength calibration step, we have a mapping between the pixel
location on the detector plane and its position in the slit mask and
wavelength: $(X_{\rm{pix}},Y_{\rm{pix}}) = D(X_{\lambda, \rm{mask}},
Y_{\rm{mask}})$.  This mapping allows us to take the flux from every
detector pixel on flat fielded frames inside the traces of the science target slits
(they are assumed to be equally wide), masking
the regions where science targets are expected to land in the corresponding
slits, to obtain a dense irregular sampling of the $(\lambda,
X_{\rm{mask}}, Y_{\rm{mask}})$ coordinate space by the sky flux.  The sky
flux in this parameter space is expected to change rapidly as a function
wavelength on a scale of the spectral resolution (i.e.  a few \AA\ in the
case of MMIRS) and slowly as a function of the mask position, because this
change is due to large scale flat fielding uncertainties and spatial
variations of the airglow emission.  Hence, the night sky emission can be
parametrized by a smoothing spline (e.g.  $b$-spline) in the wavelength
direction having about 700 nodes with the coefficients slowly varying as
functions of $X_{\rm{mask}}$ and $Y_{\rm{mask}}$, e.g.  represented by the
3rd order Legendre polynomials in two dimensions.  The number of
coefficients of the parametrization to be evaluated will be on the order of
10000--20000 depending on the chosen number of nodes and degrees of the
Legendre polynomials.

For a typical situation of a well designed slit mask, the spectral traces
occupy about 75\%\ of the detector area that is about $3\times10^6$ pixels. 
If we exclude the expected traces of unresolved science targets, this number
will decrease to about $2\times10^6$ pixels, that still provides a factor of
100--200 redundancy for the reliable computation of the $b$-spline
parametrization with the iterative outlier rejection.

\begin{figure*}
\begin{center}
\includegraphics[width=\hsize]{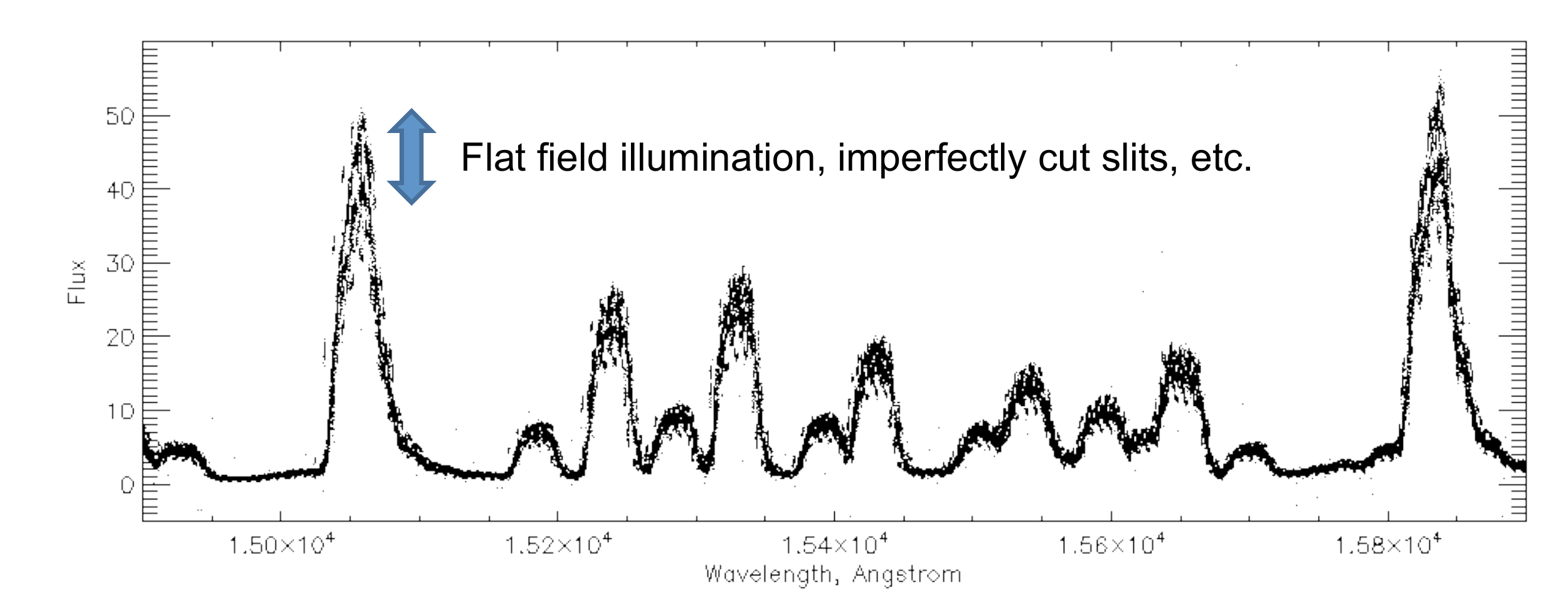}\\
\includegraphics[width=0.8\hsize]{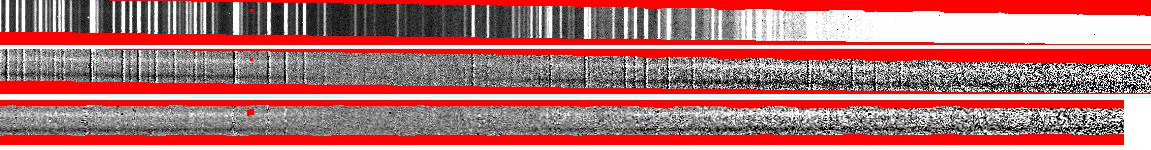}
\end{center}
\caption{An example of the sky subtraction process in the MMIRS data
reduction pipeline. The top panel
displays the oversampled night sky vector computed from the original
spectrum (it is better than a difference spectrum for illustrative purposes
only), where the variance of counts at every wavelength originates from the 
imperfect flat field illumination and/or imperfectly cut slitlets in the
mask. It is taken into accout by introducing smooth variations of the sky
model in two dimensions across the field of view (see the text).
The bottom panel shows three extracted 2D spectra from a
MOS dataset: (a) an original spectrum; (b) a difference spectrum with 
notable residuals of sky lines with clearly visible positive and negative 
traces of a science target; (c) the same difference spectrum after
the sky subtraction using the modified Kelson technique. 
\label{mmirssky}}
\end{figure*}

We use the publicly available code for the $b$-spline computation from the
{\sc sdss idlutils} package, and modify the routines in order to fit the
additional dimension with Legendre polynomials. The modified routines are
included in the pipeline distribution along with the original
two-dimensional $b$-spline fitting code.

The night sky modelling and evaluation is the longest step in the data
reduction, and it may take up-to a minute of single core CPU time on a modern
PC (3~GHz CPU).

Remarkably, our sky subtraction applied to difference frames reaches
the Poisson photon noise limits, as we demonstrate below.

\subsection{Final cosmic ray cleaning and linearisation}

After completing the sky subtraction on non-resampled data, we perform 
cosmic ray and hot pixel cleaning using the modified Laplacian
filtering technique (van Dokkum 2001) on difference images. We modify the
original code so that it can handle both negative and positive ``hits'' on
images because we are dealing with difference images.

Next, we linearize the spectra in wavelength, and rectify them (i.e.  take
out the geometric distortions from slit imaging) using the distortion map
and wavelength solutions derived at previous steps.

\subsection{Telluric correction}

Correction for telluric absorption is another important step of NIR data
reduction.  Beside \emph{OH} emission lines, there are strong water vapor
absorption bands located all over the NIR domain which vary as a function of
water vapor and airmass at the time of observation.  These bands are
responsible for nearly complete absorption of light between boundaries of
$J$ and $H$, and $H$ and $K$ photometric bands.  Even between these regions
of complete absorption, there are features reaching 15~per~cent absorption
which may look like spectral lines/bands in the target spectra if left
uncorrected.

In order to correct for the telluric absorption \citep{VCR03}, we observe
telluric standard stars which are typically A0V stars whose spectra contain
very few intrinsic features and can be relatively well modelled.  We then
derive the transmission curve by comparing the observed specrum to the model
predictions.

It is worth mentioning that the transmission curve will also include the
spectral response function of the detector, and therefore, the telluric
correction automatically performs a relative flux calibration of the data.

The standard approach is to observe telluric standard stars at different
airmasses (usually airmass values bracketing a science observation), which
allows one to compute the approximate correction of telluric absorption
bands by interpolating the correction functions.  In the regions of
intermediate and very strong absorption, however, certain absorption lines
get to full saturation and therefore the response cannot be interpolated
using only two observations at minimal and maximal airmasses.  Hence, it is
impossible to get reliable correction in these spectral regions.  It is,
however, rarely required by science programmes because the efficiency of
observations in highly absorbed regions is very low.  At the same time, in
the regions of intermediate absorption, this standard approach can
lead to significant artefacts.

\begin{figure}
\includegraphics[width=\hsize]{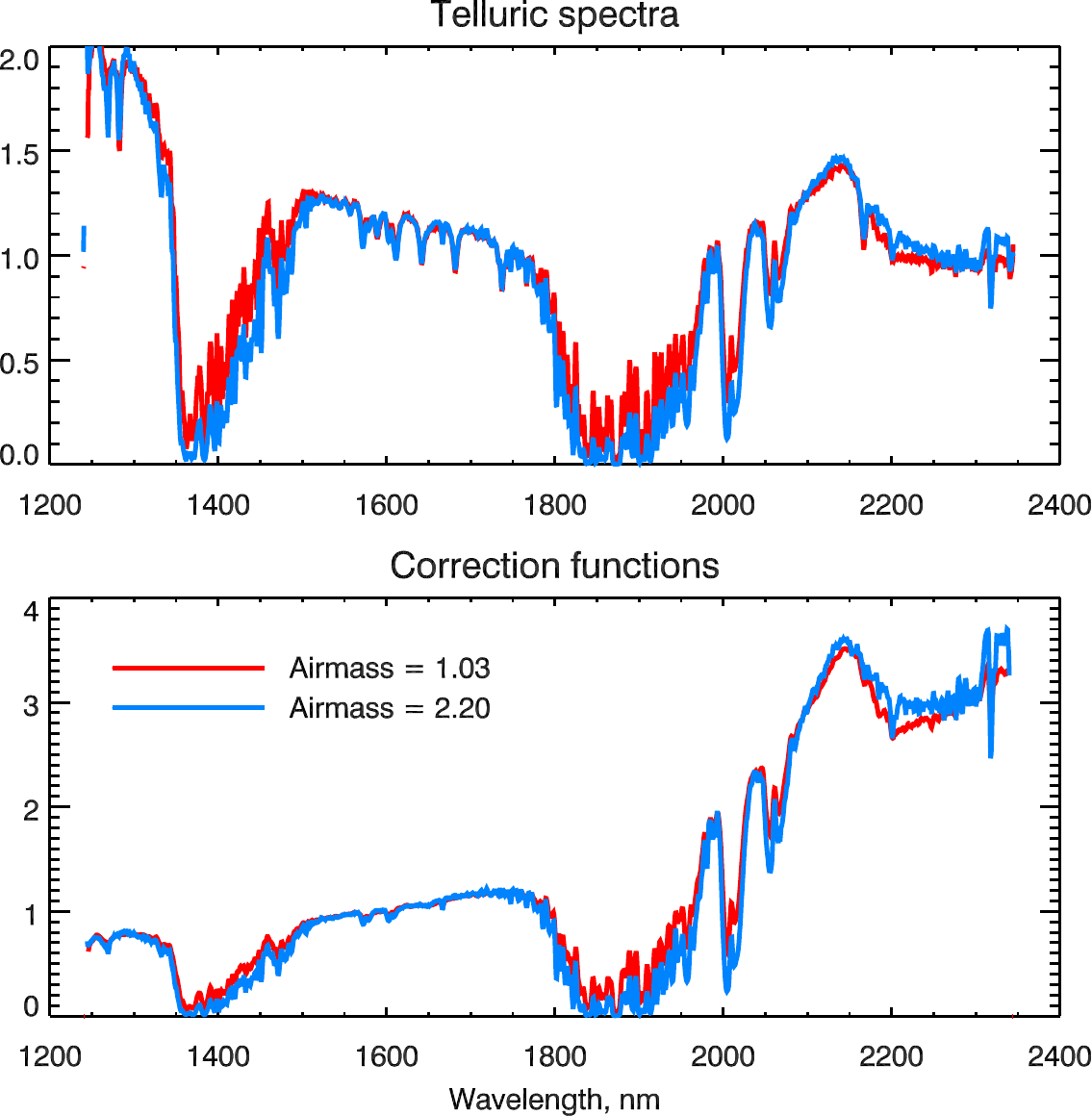}
\caption{An example of two spectra of a telluric standard star having the A0V 
spectral type obtained at different airmasses (top panel)
and the derived multiplicative correcting functions (bottom panel).\label{figtel}}
\end{figure}

An alternative solution is to use a grid of numerical models of the
atmosphere transmission, and match them against an observed spectrum
\citep{Kausch+14} assuming that it is featureless. This approach will result in
artefacts if the observed target has spectral features
where an observed science spectrum is fit against the models. It is also
very problematic to correct very faint spectra where no significant
continuum flux presents. The latter issue makes it impossible to use the fully
model-based approach to correct NIR observations of distant galaxies where
only a few emission lines are detected in a spectrum after hours of
observations.

In the MMIRS pipeline we propose a hybrid approach. We use one observation
of a telluric standard star at an airmass close to that of the science
observation, and compute the empirical atmosphere transmission function by
taking the ratio of the observed telluric spectrum and a synthetic stellar atmosphere of
that star.  This transmission function
is then fit against a grid of pre-computed theoretical models of NIR
atmospheric transmission for a given airmass, and
convolved with the MMIRS instrumental line-spread function. From this fit we derive
the best-fitting value of water vapor.  Then, we take a grid of atmospheric
transmission models at different airmass values for that value of water
vapor, and apply a differential multiplicative correction to the empirical
transmission function computed at the first step to account for the
airmass difference between the observations of the telluric standard and the
science target. A transmission function is derived in this fashion for every
science exposure and then applied to sky subtracted linearized spectra. 

Another important aspect of the telluric correction is related to handling
different slit widths.  To compute the transmission function one needs to
transform the spectrum of a high-resolution model of a A0V star into one
comparable to the observed spectrum.  This transformation depends on the
relation between the atmosphere seeing quality and the slit width.  We use
the following convention: if the atmosphere seeing quality derived from the
Gaussian FWHM of the telluric star spectrum is better than the slit width,
then we use the empirically obtained slit profile (a Gaussian with
FWHM$=$2~pix convolved with the $\Pi$-shaped profile) to convolve the model;
alternatively, if the seeing quality is worse than the slit width, we use a
simple Gaussian with FWHM equals to the seeing quality.

\subsection{Final data product}

For every observed field we process all dithered spectral difference frames
as described above to yield telluric corrected relatively flux calibrated,
two-dimensional spectra.  Then we co-add the 2D spectra of each target and
extract one-dimensional spectra using either a double-box (``positive'' and
``negative'' in order to work with the difference image) or optimal
extraction with the double-Gaussian profile (also ``positive'' and
``negative'') corresponding to the profile widths of alignment stars.  The
two-dimensional spectra from difference images are collapsed by
inverting and offsetting a half of each spectrum and then adding it to
another half.

The final data product is available in several formats: (a) multi-extension
FITS file with one extension per 2D extracted calibrated co-added spectrum;
(b) single-extension FITS file with all extracted 1D spectra; (c) a
Euro3D-FITS file \citep{KisslerPatig+04} for 2D extracted spectra; (d) a
Euro3D-FITS file for 1D extracted spectra; (e) a ESO3D-FITS file
\citep{ESO3D} for 2D extracted spectra; (f) a ESO3D-FITS file for 1D
extracted spectra.  The first two files are accompanied with files in the
same format providing flux uncertainties, while this information is stored
in the Euro3D-FITS and ESO3D formats.  MMIRS MOS spectra in the Euro3D-FITS
and ESO3D formats contain metadata making them Virtual Observatory compliant
\citep{CBLM06} and easy to visualise in specialised tools
\citep{Chilingarian+08b}.

\subsection{An example of a fully reduced dataset}

Here we provide an example of a fully reduced MOS dataset obtained in the
\emph{HK} spectral setup with 1~arcsec wide slits in December of 2011 with
the 6.5~m Magellan Clay telescope.  The slit mask was exposed for a total
of 7h~20min over two nights.  The program was targeting intermediate
redshift star-forming galaxies and active galactic nuclei and the dataset
was kindly provided for illustrative purposes by the program PIs, R.~Leiton
and E.~Daddi.

\begin{figure*}
\begin{center}
\includegraphics[width=\hsize]{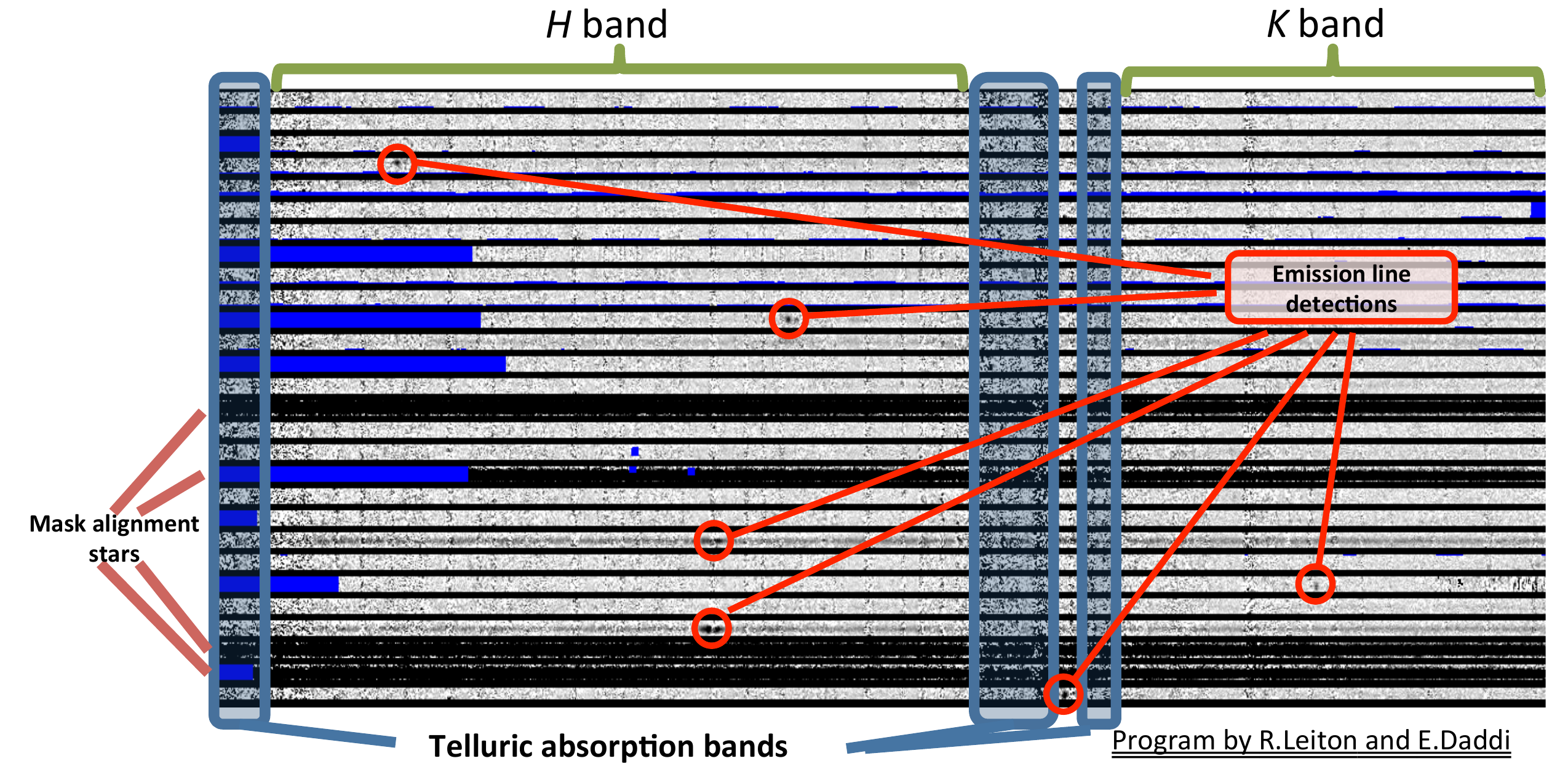}\\
\includegraphics[width=0.9\hsize]{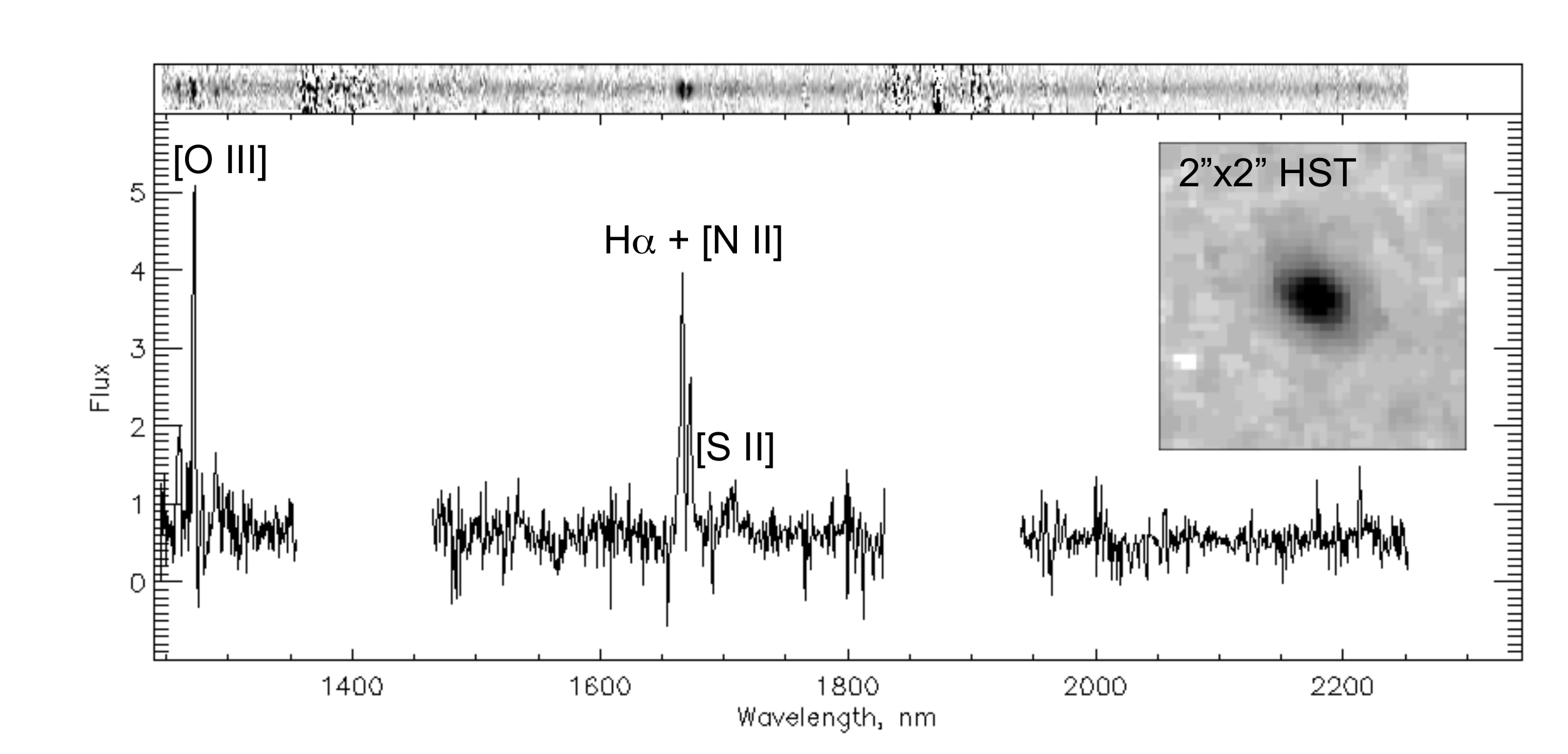}
\end{center}
\caption{Fully reduced extracted 2D slits (top panel) and an extracted 1D
spectrum (bottom panel) for a deep NIR spectroscopic dataset obtained with 
the 6.5~m Magellan Clay telescope in December of 2011. The inset in the
bottom panel displays a 2$\times$2~arcsec fragment of the Hubble Space
Telescope image of a $z=1.54$ galaxy having the total Vega \emph{H} band 
magnitude $m_H = 19.8$~mag. The integration time was 7h~20min with the
0.6~arcsec average seeing quality.
\label{MMIRSdataset}}
\end{figure*}

In Fig~\ref{MMIRSdataset} we display stacked 2D, sky subtracted, telluric
corrected, rectified spectra linearized in wavelength, and a plot of an
extracted 1D spectrum for one of the objects.  Emission line detections are
clearly visible in several slits.  Residuals of the night sky \emph{OH}
emission lines are barely visible demonstrating the high quality of sky
subtraction by our data reduction pipeline.

\section{QUALITY OF THE SKY SUBTRACTION}
\label{sect:skysub}

\begin{figure*}
\includegraphics[width=\hsize]{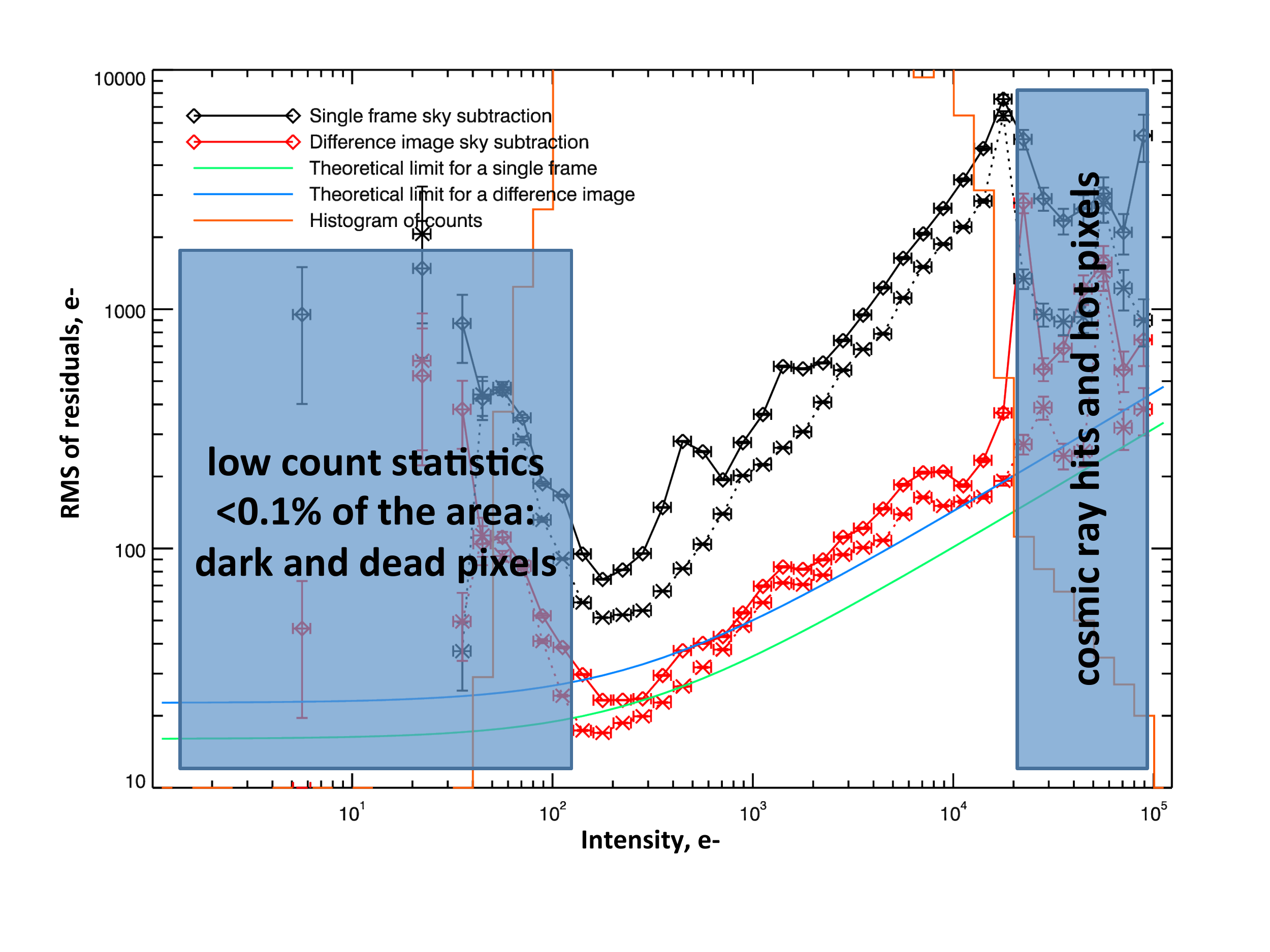}
\caption{The assessment of the sky subtraction quality provided by the MMIRS
pipeline for an observation using the $HK$ grism with the pre-2014B version of the
instrument equipped with the HAWAII-2 detector. The blue solid line corresponds
to the theoretical sky subtraction quality limit from the Poisson photon
statistics and read-out noise for a double correlated read. The solid red
curve displays the measured variance of the sky residuals as a function of
intensity. The dashed red curve presents the same measurements made using 
$\sigma$-clipping in the distributions of counts in order to eliminate the
effects of residual cosmic ray hits and warm or cold pixels.
\label{figSkySubQ}}
\end{figure*}

In order to assess the quality of the sky subtraction, we compute the
histogram of counts (converted to electrons) on a raw single spectral frame
binned by intensity.  We then compute the standard deviations of counts in
every intensity bin with 3-$\sigma$ outlier rejection in order to remove the
effects of uncorrected hot pixels and cosmic ray hits.  If a bin in
intensity is sufficiently narrow, the standard deviation of counts on a sky
subtracted frame corresponding to the narrow range of intensities on a non
sky subtracted frame will provide a diagnostics of the sky subtraction
quality.  This computation is repeated for a single spectral frame and a
difference spectrum.  The theoretical noise limit corresponds to the square
root of intensity (Poisson limit) at intermediate and high count rates, and
approaches the plateau set by the read-out noise at low count rates.  For
the case of a difference frame, the read-out noise plateau is higher by a
factor of $\sqrt 2$.  In Fig.~\ref{figSkySubQ} we present the result of this
computation for a typical MMIRS MOS observation with the exposure time of
300~sec.

From this plot it is clear that our sky subtraction approach applied to
difference spectra leaves residuals in \emph{OH} lines consistent with the
Poisson photon statistics.

However, when applied to single spectral exposures, it leaves significant
residuals exceeding the Poisson limit by a factor of up-to 10.  The main
reason for this is probably the scattered light in the spectrograph, given
that the pixel-to-pixel sensitivity variations of the {\sc hawaii-2} detector 
are wavelength dependent.  The dispersed light in the spectral traces on a detector
plane is nearly monochromatic in every pixel, while the diffuse component of
the scattered light is ``white''.  The diffuse component has the intensity
of about 2\%, while the pixel-to-pixel sensitivity variations reach 40\%\ 
and a change by a factor of several as a function of wavelength.  Therefore, on
a single exposure, up-to 1\%\ of the residual
intensity will appear as noise, an increase of an order of magnitude in the
uncertainties over the Poisson photon statistics.  When
dealing with difference spectra, on the other hand, the diffuse light modulated by
pixel-to-pixel sensitivity variations almost completely cancels itself
because of the subtraction.  Hence, the sky model becomes representative of
the residual \emph{OH} lines and the sky subtraction becomes Poisson
limited.

%We expect that after the detector upgrade the situation will improve.
%According to the specifications, {\sc hawaii-2rg} detectors have
%pixel-to-pixel sensitivity variations under 10\%. Even if the colour terms
%of the response are as strong as in the case of {\sc hawaii-2}, this will
%improve the sky subtraction quality on a single frame by a factor of 4
%already approaching the Poisson limit.

\section{SUMMARY}
\label{sect:summary}

We present a spectroscopic data reduction pipeline for the multi-object 
near-infrared spectrograph MMIRS capable of reducing large data volumes in 
an automatic way. We describe the observational protocol that should be 
followed by observers in order to make the data compatible with our 
pipeline.

We implement the pipeline as a stand-alone software package in {\sc idl} 
with minimal software dependencies.  Our pipeline supports both the 
long-slit and multi-object spectroscopic modes.  Pipeline execution is 
controlled by text parameter files with the format resembling that of FITS 
headers.  The text files can be created manually, or generated using a 
script that accesses a relational database containing the observational 
metadata.

We implement novel algorithms for sky subtraction and telluric correction.  
We demonstrate that our sky subtraction technique provides sky subtraction 
quality close to the Poisson limit set by the photon statistics.  Our 
telluric correction technique requires only one telluric standard 
observation, and differentially adjusts the standard to the airmass of a 
corresponding science observation using a grid of synthetic atmosphere 
transmission models.

The pipeline can be used in real-time at a telescope to perform quick data
reduction in order to assess data quality.

With minor modifications our pipeline can be adopted as a data reduction 
solution for other long-slit and multi-object optical and NIR spectrographs 
operated by different observatories.  For example, we anticipate that this 
pipeline will become the workhorse data reduction pipeline for Binspec, a 
multi-slit optical spectrograph that will soon be commissioned on the MMT.

%%%%%%%%%%%%%%%%%%%%%%%%%%%%%%%%%%%%%%%%%%%%%%%%%%%%%%%%%%%%%
\section{ACKNOWLEDGMENTS}

The project is carried out at the Telescope Data Center supported by the
Smithsonian Astrophysical Observatory.  We acknowledge the support of MMIRS
operations and instrument software by Maureen Conroy, Anne Matthews, and
John Roll of the Smithsonian Astrophysical Observatory, and support of
observations by Dave Osip, Povilas Palunas, and technical staff of the Las
Campanas Observatory in Chile.  We thank our colleagues Jessica Mink and
Andrew Szentgyorgyi for useful discussions.  IC acknowledges the additional
support from the Russian Science Foundation project 14-22-00041.

%%%%%%%%%%%%%%%%%%%%%%%%%%%%%%%%%%%%%%%%%%%%%%%%%%%%%%%%%%%%%
%%%%% References %%%%%

%\begin{thebibliography}{}
%
%\end{thebibliography}

\bibliographystyle{apj}
\bibliography{mmirs_pipeline}

\begin{thebibliography}{}
\expandafter\ifx\csname natexlab\endcsname\relax\def\natexlab#1{#1}\fi

\bibitem[{{Afanasiev} {et~al.}(2001){Afanasiev}, {Dodonov}, \&
  {Moiseev}}]{ADM01}
{Afanasiev}, V.~L., {Dodonov}, S.~N., \& {Moiseev}, A.~V. 2001, in Stellar
  Dynamics: from Classic to Modern, ed. L.~P. {Ossipkov} \& I.~I. {Nikiforov},
  103

\bibitem[{{Bertin}(2001)}]{Bertin01}
{Bertin}, E. 2001, in Mining the Sky, ed. A.~J. {Banday}, S.~{Zaroubi}, \&
  M.~{Bartelmann}, 353

\bibitem[{{Castelli} \& {Kurucz}(2004)}]{CK04}
{Castelli}, F., \& {Kurucz}, R.~L. 2004, ArXiv astro-ph/0405087

\bibitem[{{Chilingarian} {et~al.}(2006){Chilingarian}, {Bonnarel}, {Louys}, \&
  {McDowell}}]{CBLM06}
{Chilingarian}, I., {Bonnarel}, F., {Louys}, M., \& {McDowell}, J. 2006, in
  Astronomical Society of the Pacific Conference Series, Vol. 351, Astronomical
  Data Analysis Software and Systems XV, ed. C.~{Gabriel}, C.~{Arviset},
  D.~{Ponz}, \& S.~{Enrique}, 371

\bibitem[{{Chilingarian} {et~al.}(2008){Chilingarian}, {Bonnarel}, {Louys},
  {Zolotukhin}, {Royer}, {Jegouzo}, {Le Sidaner}, {Fernique}, \&
  {Boch}}]{Chilingarian+08b}
{Chilingarian}, I., {Bonnarel}, F., {Louys}, M., {et~al.} 2008, in Astronomical
  Spectroscopy and Virtual Observatory, ed. {M.~Guainazzi \& P.~Osuna}, 125--+

\bibitem[{{Chilingarian} \& {Bergond}(2010)}]{CB10}
{Chilingarian}, I.~V., \& {Bergond}, G. 2010, \mnras, 405, L11

\bibitem[{{Chilingarian} {et~al.}(2009){Chilingarian}, {De Rijcke}, \&
  {Buyle}}]{CdRB09}
{Chilingarian}, I.~V., {De Rijcke}, S., \& {Buyle}, P. 2009, \apjl, 697, L111

\bibitem[{{Chilingarian} {et~al.}(2007){Chilingarian}, {Prugniel},
  {Sil'chenko}, \& {Afanasiev}}]{CPSA07}
{Chilingarian}, I.~V., {Prugniel}, P., {Sil'chenko}, O.~K., \& {Afanasiev},
  V.~L. 2007, \mnras, 376, 1033

\bibitem[{{Davies} {et~al.}(2013){Davies}, {Agudo Berbel}, {Wiezorrek},
  {Cirasuolo}, {F{\"o}rster Schreiber}, {Jung}, {Muschielok}, {Ott}, {Ramsay},
  {Schlichter}, {Sharples}, \& {Wegner}}]{Davies+13}
{Davies}, R.~I., {Agudo Berbel}, A., {Wiezorrek}, E., {et~al.} 2013, \aap, 558,
  A56

\bibitem[{{Kausch} {et~al.}(2014){Kausch}, {Noll}, {Smette}, {Kimeswenger},
  {Horst}, {Sana}, {Jones}, {Barden}, {Szyszka}, \& {Vinther}}]{Kausch+14}
{Kausch}, W., {Noll}, S., {Smette}, A., {et~al.} 2014, in Astronomical Society
  of the Pacific Conference Series, Vol. 485, Astronomical Data Anaylsis
  Softward and Systems XXIII, ed. N.~{Manset} \& P.~{Forshay}, 403

\bibitem[{{Kelson}(2003)}]{Kelson03}
{Kelson}, D.~D. 2003, \pasp, 115, 688

\bibitem[{{Kissler-Patig} {et~al.}(2004){Kissler-Patig}, {Copin}, {Ferruit},
  {P{\'e}contal-Rousset}, \& {Roth}}]{KisslerPatig+04}
{Kissler-Patig}, M., {Copin}, Y., {Ferruit}, P., {P{\'e}contal-Rousset}, A., \&
  {Roth}, M.~M. 2004, Astronomische Nachrichten, 325, 159

\bibitem[{{K\"ummel} {et~al.}(2012){K\"ummel}, {Ballester}, \&
  {Kuntschner}}]{ESO3D}
{K\"ummel}, M., {Ballester}, P., \& {Kuntschner}, H. 2012, in ESO Very Large
  Telescope Technical Report, VLT--SPE--ESO--19500--5667,
  \url{ftp://ftp.eso.org/pub/dfs/pipelines/doc/VLT-SPE-ESO-19500-5667_DataFormat.pdf}

\bibitem[{{Kurucz}(2005)}]{Kurucz05}
{Kurucz}, R.~L. 2005, Memorie della Societa Astronomica Italiana Supplementi,
  8, 14

\bibitem[{{McLeod} {et~al.}(2012){McLeod}, {Fabricant}, {Nystrom}, {McCracken},
  {Amato}, {Bergner}, {Brown}, {Burke}, {Chilingarian}, {Conroy}, {Curley},
  {Furesz}, {Geary}, {Hertz}, {Holwell}, {Matthews}, {Norton}, {Park}, {Roll},
  {Zajac}, {Epps}, \& {Martini}}]{McLeod+12}
{McLeod}, B., {Fabricant}, D., {Nystrom}, G., {et~al.} 2012, \pasp, 124, 1318

\bibitem[{{Newman} {et~al.}(2013){Newman}, {Cooper}, {Davis}, {Faber}, {Coil},
  {Guhathakurta}, {Koo}, {Phillips}, {Conroy}, {Dutton}, {Finkbeiner}, {Gerke},
  {Rosario}, {Weiner}, {Willmer}, {Yan}, {Harker}, {Kassin}, {Konidaris},
  {Lai}, {Madgwick}, {Noeske}, {Wirth}, {Connolly}, {Kaiser}, {Kirby},
  {Lemaux}, {Lin}, {Lotz}, {Luppino}, {Marinoni}, {Matthews}, {Metevier}, \&
  {Schiavon}}]{Newman+13}
{Newman}, J.~A., {Cooper}, M.~C., {Davis}, M., {et~al.} 2013, \apjs, 208, 5

\bibitem[{{Vacca} {et~al.}(2003){Vacca}, {Cushing}, \& {Rayner}}]{VCR03}
{Vacca}, W.~D., {Cushing}, M.~C., \& {Rayner}, J.~T. 2003, \pasp, 115, 389

\bibitem[{{Vacca} {et~al.}(2004){Vacca}, {Cushing}, \& {Rayner}}]{VCR04}
---. 2004, \pasp, 116, 352

\bibitem[{{van Dokkum}(2001)}]{vanDokkum01}
{van Dokkum}, P.~G. 2001, \pasp, 113, 1420

\end{thebibliography}

%\clearpage

\appendix

\section{TECHNICAL DETAILS ABOUT THE PIPELINE CODE}

\subsection{The pipeline calibration data}

The pipeline is distributed with a set of calibration data required for the
proper functionality of the code located in the \emph{calib\_MMIRS/}
directory.  They are organized in a set of sub-directories corresponding to
their functional purposes.

The detector specific calibrations stored in the \emph{calib\_MMIRS/H2\_56/}
folder include the specific information for the HAWAII-2 detector serial
number 56 that had been used in MMIRS in 2009-2014.  There are two files
presenting the detector layout for the read-out modes using 4 and 32
amplifiers.  They have a format of the binary FITS table with one row per
amplifier defining its minimal and maximal pixel coordinates in X and Y
axes, and the read-out direction in the detector, ``1'' for the horizontal
and ``2'' for the vertical directions respectively.  The
\emph{nonlinearity/} sub-folder contains the definition of the correcting
polynomial that needs to be applied to the raw counts in order to put them
into the linear scale.  It also contains the bias and saturation values. 
This information is provided per each row or column (depending on the
read-out direction) for every amplifier.  Hence, 4096 sets of values are
given as the entries of the binary FITS table.  Below we describe the
algorithm of the derivation of the non-linearity correction.

The \emph{calib\_MMIRS/LS/} and \emph{calib\_MMIRS/MOS/} directories contain
the template distortion maps for all grism/filter combinations in the
longslit and MOS spectral configurations respectively.  The longlist folder
also contains the mask definition files (one for each slit width) describing
the long slit layout in the format identical to that of the MOS masks. These
files are replicated into the data reduction directories as
``mask\_mos.txt'' in order to use the same data reduction routines as
those for the MOS mode without introducing any modifications to the code.

The \emph{calib\_MMIRS/linelists/} folder contains lists of spectral
lines with wavelengths and relative intensities used for the wavelength
calibration of MMIRS spectra (see above). There are two lists, one
containing argon lines from the arc line lamp, and another one with the {\sc
oh} lines when the wavelength solution is computed from the science frames
themselves.

The information required to perform the telluric correction is stored in the
\emph{calib\_MMIRS/sky\_transmission/} and \emph{calib\_MMIRS/telluric/}
directories.  The former folder contains the grid of amosphere transmission
models computed with the {\sc atran} code for the Gemini-South telescope and
downloaded from the Gemini observatory
web-site\footnote{http://www.gemini.edu/}.  The latter directory contains
high resolution {\sc atlas9} models \citep{CK04,Kurucz05} of synthetic stellar
atmospheres for the \emph{A0V} and \emph{G0V} spectral types corresponding
to stars used as telluric standards during MMIRS spectroscopic observations. 
The Gemini-South telescope is geographically located relatively nearby
(about 150~km) to the Las Campanas Observatory in Chile where one would
expect similar atmosphere conditions, therefore we did not compute the
specific atmosphere transmission models for LCO but simply used those
provided by Gemini.

\subsubsection{Defining the non-linearity correction}

The non-linearity correction is a crucial step in the primary data reduction
because NIR detectors are known to have a non-linear response at all
intensity levels \citep{VCR04}.  In addition to the non-linearity, the
detector has a so-called reset effect, that is an additive constant signal
level that appears after the detector reset (cleaning) procedure at the
beginning of each exposure and then declines exponentially with time.  A
natural way to calibrate the non-linear response is to take a long
up-the-ramp exposure of a constant source of light such as imaging flat
field until the detector saturates.  The intensity of the source, however,
has to be relatively low to provide enough sampling of the response curve. 
Then, if the detector response is ideally linear, the counts will grow
linearly in time until the saturation and any measured deviation from the
linear behavior will be due to the detector reponse.

Since for practical reasons it is nearly impossible to have an absolute
calibration of the flat field light source, we either have to deal with
differential correcting functions or define a count range where the
non-linearity is the lowest and cross-calibrate the bright and faint tails
of the response curve to that region.  We decided to use the latter option
and found that the lowest non-linearity of the MMIRS HAWAII-2 detector is
achieved between 5000 and 9000 counts above the bias level.

The internal flat field lamp in the imaging mode is too bright to obtain
good sampling of the response curve, therefore we used science observations
of stellar fields in different photometric bands that we found in the MMIRS
archive.  Thus, the night sky played a role of the constant light source. 
We manually chose exposures without extended sources and avoided crowded
fields.  The count rates (counts per second) from the night sky 
vary by two orders of magnitude in different filters ($Y$ being the
faintest and $H$ being the brightest). This allowed us to sample well the
detector response curve from very low fluxes (100 counts) up to the
saturation levels (40000 counts).

We started with the $K$-band filter observations and defined the reponse
curve at the 5000--10000 count range, then used $Y$ and $J$ band
observations in the low end and $H$ and $HK$ band observations in the high
end cross-calibrating them to the initial part of the curve. This operation
was repeated for different up-the-ramp settings. Then, we performed the
fitting of the response curve for every row/column in every read-out amplifier
(see above the description of the amplifier layout).

We fit the reset-effect by an exponential function and find an
exponential decay timescale of 2.7~sec.  Therefore, it has very little
effect on the data obtained in the up-the-ramp settings with read-out
cadence of 5~sec or longer.  This additive effect is taken out during the
primary reduction process separately from the non-linearity correction.

Then we fit the non-linear reponse curve in the 100--39000 count range by
the 9th order polynomial function and obtained the RMS residuals of less
than 2 counts in the whole range of intensities.  The response of the
detector is non-linear at a few per~cent level under 3000 and above 15000
counts, however after applying the correction we
reach a sub-per~cent level of non-linearity.

%\begin{figure}
%%\includegraphics{nonlinfig.ps}
%\caption{The non-linear response curve of the HAWAII-2 detector measured
%from the night sky observations in different photometric bands.\label{fignonlin}}
%\end{figure}

\subsubsection{Obtaining the initial wavelength solution approximations}

A key to the efficient and precise wavelength calibration of MMIRS
spectral data is the availability of high quality (to within a couple of pixels)
initial approximation of the wavelength solution, sufficient for easy automatic
identification of spectral lines. The initial approximations for all MMIRS
grism settings are kept inside the pipeline code. Here we briefly describe
how we obtain them in order to show how we will handle new grisms. This
procedure must be done for every new grism installed in the
instrument, but does not need to be performed again by pipeline users.

From longslit spectroscopic observations we measure the curvature of the
slit image defining the average pixel shift of the wavelength solution with
respect to the slit center.  This offset is parametrized by the 3rd order
polynomial of the Y coordinate that will also apply to MOS slit masks.

Then we manually identify a dozen arc lines in several MOS slits covering a
large range of X slit mask positions, apply the offsets corresponding to the
Y positions computed at the previous step, and then fit dispersion relations
(i.e.  wavelength as a function of pixel position on a detector plane) with
low order one-dimensional polynomials.  Finally, we fit the behavior of the
zeroth and first order coefficients as a function of X slit mask positions. 
The approximate solutions are stored in the pipeline code.

\subsection{Dependencies and modified versions of third-party
software used in the pipeline}

The current version of the pipeline relies on one third-party {\sc idl}
package and a few routines extracted from their original packages and
modified for our purposes. Here we briefly describe these routines and the
ways they are used in the pipeline.

The only package that must be installed in the system before the
pipeline can be used is the IDL Astronomy User's
Library\footnote{\url{http://idlastro.gsfc.nasa.gov/}}, or {\sc astrolib},
distributed by NASA Goddard Space Flight Center.  We use a number of
routines from {\sc astrolib} to perform the input-output operations on FITS
files and binary tables, {\sc idl} data structures, and also some routines
to compute outlier resistant statistics on datasets.

Here we present a list of modified and/or extracted routines used in the
pipeline.

\begin{itemize}
\item {\sc find\_1d.pro, cntrd\_1d.pro, gcntrd\_1d.pro} are modified
versions of the corresponding routines from the {\sc idlphot} sub-package in
{\sc astrolib} which were modified in order to operate on one-dimensional
arrays (vectors in the {\sc idl} terminology) rather than on two-dimensional
images

\item {\sc sfit\_2deg.pro} is a rewritten version of the {\sc sfit.pro} 2D
surface fitting routine from the {\sc idl} distribution that allows one to
use different degrees of the polynomial approximation on two spatial
dimensions

\item {\sc goodpoly\_err.pro} is a modified version of the {\sc goodpoly.pro}
outlier resistant polynomial fitting routine by Marc
Buie\footnote{\url{http://www.boulder.swri.edu/~buie/}} that allows one to specify
measurement errors

\item {\sc la\_cosmic\_array.pro} is a modified version of the Laplacian
cosmic ray filtering technique \citep{vanDokkum01} that allows us to deal
with data arrays instead of lists of input images and also has a keyword to
reject negative outliers that is required to deal with difference spectral
frames

\item Extracted routines from SDSS {\sc
idlutils}\footnote{\url{http://www.sdss3.org/dr8/software/idlutils.php}}: we use
certain mathematical algorithms from the {\sc idlutils} package and their
dependencies including the $b$-spline fitting routines. We also provide
modified versions of the $b$-spline fitting code with the ``\_3d'' suffix
that allow one to fit a $b$-spline approximation on a scattered dataset
with two extra dimensions used in the MMIRS sky subtraction algorithm. 

\end{itemize}

\section{EXAMPLE OF A TASK CONTROL FILE}

Here we provide an example of a MMIRS pipeline task control file generated
automatically by a {\sc perl} script using a {\sc PostgreSQL} database
containing metadata of all MMIRS observations.

\begin{verbatim}
SIMPLE  =                                 T / FITS-like header
LONGSTRN= 'OGIP 1.0'           / The OGIP long string convention may be used.
RAW_DIR = '/data/mmirs/Archive/rawdata/data/MMIRS/2011/2011.1203/'
R_DIR   = '/d1/data1/mmirs/preproc/2011/2011.1203/'
W_DIR   = '/d1/data1/mmirs/reduced/2011/2011.1203/lei1.a1_mask1_HK_HK/01/'
RAWEXT  = '.gz'
INSTRUME= 'MMIRS' / spectrograph name
SLIT    = 'mos' / slit name or MOS
GRISM   = 'HK' / grism
FILTER  = 'HK' / filter
BRIGHT  = 0 / bright source 1=Yes 0=No
SCI     = 'lei1.a1_mos.8456' / science frame (only one)
SCI2    = 'lei1.a1_mos.8457' / second science frame (only one)
DITHPOS = 1.8000 / current dithering position in arcsec
DITHPOS2= -1.4000 / dithering position for the second frame
DARKSCI = 'dark.8653,dark.8654,dark.8655,dark.8656,dark.8657' / dark frame(s) for science
ARC     = 'comp_mos.8468' / arc frames
DARKARC = 'dark.8638,dark.8639,dark.8640,dark.8641,dark.8642' / dark frame(s) for arc
FLAT    = 'flat_mos.8470'
DARKFLAT= 'dark.8628,dark.8629,dark.8630,dark.8631,dark.8632' / dark frame(s) for flat
STAR01  = 'HIP-16904_mos.8491,HIP-16904_mos.8492,HIP-16904_mos.8494,&'
CONTINUE  'HIP-16904_mos.8514,HIP-16904_mos.8516'
DARKST01= 'dark.8638,dark.8639,dark.8640,dark.8641,dark.8642'
STTYPE01= 'a0v'
STAR02  = 'HIP-16904_mos.8596,HIP-16904_mos.8597,HIP-16904_mos.8598,&'
CONTINUE  'HIP-16904_mos.8599,HIP-16904_mos.8600'
DARKST02= 'dark.8638,dark.8639,dark.8640,dark.8641,dark.8642'
STTYPE02= 'a0v'
COMMENT   flat fielding settings
FFSCFLAT= 0  / subtract scattered light from flat field?
FFSCSCI = 0  / subtract scattered light from science frames?
FFNSLIT = -1 / normalize to slit #N, -1 for per-slit normalization
COMMENT   wavelength calibration settings
WLNDEG  = 3 / polynomial degree along dispersion
WLYNDEG = 3 / polynomial degree across dispersion
WLDEBUG = 0 / debug setting
WLPLOT  = 0 / plot the arc line identification spectrum on the screen
WLADJ   = 1 / adjust wavelength solution using all slits
WLOH    = 1 / use OH lines to build the wavelength solution
SKIPLINE= ' ' / arc lines NOT to use
COMMENT   sky subtraction settings
SSADJWL = 1 / use the adjusted wavelength solution?
SSDIM   = 3 / 3-dimension or 2-dimension sky model?
SSDEBUG = 0 / debug for sky subtraction
COMMENT   linearisation settings
LINADJWL= 1 / use the adjusted wavelength solution?
LINSSBOX= 1 / sky subtraction in linearized spectra in all "BOX" slits
LINSSTRG= 1 / sky subtraction in linearized spectra in all "TARGET" slits
COMMENT   extraction settings
EXTMETH = 1 / 1 - box extraction, 2 - optimal extraction
EXTAPW  = 5 / extraction aperture in pixels, window with or gaussian FWHM
EXTDFBOX= 0 / use alignbox star positions to get empirical dithering offset?
EXTBOXEX= 'obj-sky' / image type to use alignbox positions from
COMMENT   process the following reduction steps (yes=1/no=0)
S01PROC = 1 / up-the-ramp fitting (if needed) and dark subtraction
S02PROC = 1 / distortion map creation
S03PROC = 1 / flat-fielding and 2D slit extraction
S04PROC = 1 / wavelength calibration
S05PROC = 1 / sky subtraction
S06PROC = 1 / linearisation
S07PROC = 1 / extraction
S08PROC = 1 / telluric star processing
S09PROC = 0 / telluric correction
COMMENT   create_task.pl -m lei1.a1 -t mask1 -g HK -f HK -d lei1_HK_HK/ -s HIP% 
END
\end{verbatim}

\end{document}